\documentclass[sigconf,nonacm]{acmart}

\usepackage{booktabs}
\usepackage{graphicx}
\usepackage{amsmath}
\usepackage{algorithm}
\usepackage[noEnd=true,indLines=true]{algpseudocodex}
\usepackage{array}
\usepackage{comment}
\usepackage{listings}
\usepackage{tabularx}
\usepackage{tikz}
\usetikzlibrary{arrows.meta,fit,positioning}
\newcolumntype{Y}{>{\raggedright\arraybackslash}X}

\begin{document}

\title{\SYSNAME{}: Dynamic Folding of Concurrent Analytical Queries}

\author{Genki Kimura}
\affiliation{
  \institution{The University of Tokyo}
  \city{Tokyo}
  \country{Japan}
}
\email{kimura-g@tkl.iis.u-tokyo.ac.jp}

\author{Kazuo Goda}
\affiliation{
  \institution{The University of Tokyo}
  \city{Tokyo}
  \country{Japan}
}
\email{kgoda@tkl.iis.u-tokyo.ac.jp}

\newcommand{\SYSNAME}{\textsc{GraftDB}}
\newcommand{\query}[1]{\ensuremath{Q_{#1}}}
\lstdefinestyle{sqlblock}{
  language=SQL, basicstyle=\ttfamily\small, columns=fullflexible, keepspaces=true,
  showstringspaces=false, aboveskip=0.5\baselineskip, belowskip=0.5\baselineskip, xleftmargin=1em }

\begin{abstract}

Analytical database systems serve as foundational infrastructure for
knowledge discovery across many domains.
Day after day, researchers, practitioners, and increasingly AI-driven
agents issue analytical queries, inspect their results, and refine
their inquiries.
An analytical database system thus receives and processes diverse
analytical queries that arrive over time and execute concurrently.
Such workloads can create redundant execution work across independently issued
queries.
Exploiting this overlap to optimize query processing as a whole is a
critical technical challenge.

This paper presents \SYSNAME{}, a multi-query execution engine that
dynamically folds a later-arriving query into a running execution,
reusing previously performed work
and sharing subsequently performed work.
\SYSNAME{} achieves dynamic folding with \emph{state-centric execution},
which treats operator state accumulated during execution
not as owned by a single query,
but as shared state that any compatible query can observe or contribute to.
Each query observes shared state through a \emph{per-query state lens},
which lets the query observe that state only after the relevant input has
been incorporated and receive only rows or state fragments valid under the
query's semantics.
For an arriving query, \emph{query grafting} identifies
operator state that already satisfies part of the query's
requirements and work that can still be shared to satisfy the rest.
Together, these mechanisms let \SYSNAME{} share work across overlapping
analytical queries and reduce redundant execution work.
Experiments using TPC-H-derived instances of dynamic concurrent workloads show
that \SYSNAME{} achieves up to 2.17 times higher throughput than a
same-engine isolated-execution baseline.
Under overloaded open-loop arrivals, \SYSNAME{} reduces P95 response time to
as low as 0.17 times the same baseline's P95 response time.
\end{abstract}

\maketitle

\begingroup\small\noindent\raggedright\textbf{Artifact Availability:}\\
Artifact materials are available at
\url{https://github.com/dbc-utokyoiis/GraftDB}.
\par\endgroup

\section{Introduction}

Analytical database systems support many forms of data analysis.
Human analysts, applications, and AI-driven agents independently
issue analytical queries over shared data~\cite{conf/iclr-ws/ZhangSLZ24,
conf/iclr/LeiCYCSSSGHYZX025}.
Together, these sources create \emph{dynamic concurrent workloads}, in which
analytical queries arrive over time, query
executions overlap, and the active query set changes while the system
is running~\cite{journals/pvldb/PsaroudakisAA13,journals/pvldb/WuCHN13}.

A central opportunity in such workloads is to reduce
redundant work across queries that coexist in time.
An arriving query may differ from already-running queries
in parameters, predicates, and arrival time, but it may still
require much of the same underlying computation.
If each query runs in isolation, the system repeats parts of that computation across
concurrent executions.

Prior systems reduce such redundancy in several ways.
Some systems reuse a stored result, a materialized view, or an
intermediate result saved before the query runs
\cite{conf/sigmod/GoldsteinL01,conf/sigmod/IvanovaKNG09,
conf/pods/LevyMSS95,conf/icde/NagelBV13}.
Other systems optimize a known set of queries and create shared plans
before execution
\cite{conf/sigmod/Finkelstein82,journals/pvldb/GiannikisAK12,
journals/pvldb/GiannikisMAK14,journals/pvldb/MakreshanskiGAK16,
conf/sigmod/RoySSB00,journals/tods/Sellis88,conf/sigmod/ZhouLFL07}.
Runtime-sharing systems coordinate scans, operator pipelines, or shared
joins while queries are running
\cite{conf/sigmod/ArumugamDJPP10,journals/pvldb/CandeaPV09,
conf/sigmod/HarizopoulosSA05,journals/pvldb/UnterbrunnerGAFK09,
conf/vldb/ZukowskiHNB07}.
These approaches share work through saved results, known query sets,
scans, pipelines, or join structures.

\begin{figure}[t]
  \centering
  \includegraphics[width=\linewidth]{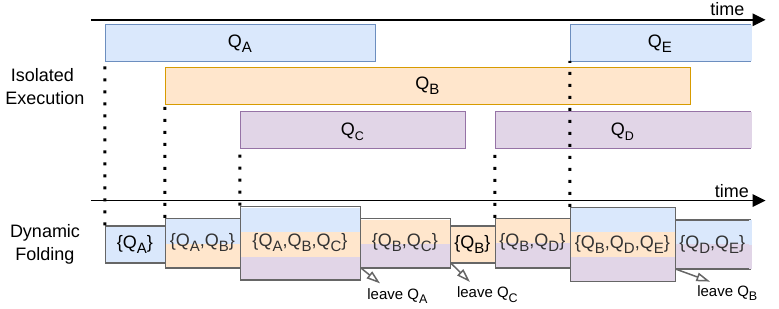}
  \caption{
    Dynamic folding integrates arriving analytical queries into an
    evolving shared execution.
    In isolated execution, overlapping queries run separately.
    In dynamic folding, the shared execution carries the current set of
    attached queries and changes that set as queries arrive and complete.
  }
  \Description{Timeline comparing isolated executions with dynamic folding.
  In isolated execution, each query runs as a separate bar over time.
  In dynamic folding, the carried query set changes at arrival and completion points, so the
  shared execution can continue as individual queries attach and complete.}
  \label{fig:intro-dynamic-attachment}
\end{figure}

Dynamic concurrent workloads create another opportunity
inside a running execution.
While a query is executing, its stateful operators accumulate operator
state such as hash tables and aggregate accumulators.
When a later query arrives, part of that operator state may already
satisfy part of the query's requirements,
and the running execution may still have work remaining that can contribute the rest
to the same state.
The challenge is to make this partially accumulated, still-changing
state safely usable across queries with different predicates,
parameters, and arrival times.

Dynamic folding is an execution strategy for dynamic concurrent workloads.
Rather than executing an arriving query in isolation,
it folds the query into a running execution:
the query reuses compatible operator state already accumulated there,
shares subsequent computation that contributes to the same state, and leaves
when its own work completes.
Figure~\ref{fig:intro-dynamic-attachment} illustrates the difference
between conventional isolated execution, where overlapping queries run separately,
and dynamic folding, where arriving queries attach to an evolving shared execution and
leave when their work completes.

We present \SYSNAME{}, a multi-query execution engine for dynamic
concurrent workloads.
\SYSNAME{} achieves dynamic folding with \emph{state-centric execution},
which treats the state of stateful operators, such as hash joins and
aggregations, not as state owned by the query that first produced it,
but as shared state that any compatible query can observe or contribute to.
Two runtime mechanisms make this practical.
A \emph{per-query state lens} lets each query observe shared state under that query's
semantics:
it determines which part of the state is complete for the query and which rows
or state fragments the query may receive.
\emph{Query grafting} attaches an arriving query to compatible shared state by
identifying both the state already present and the producer work that can still
contribute to it.
Together, per-query state lenses and query grafting let \SYSNAME{} share state across
queries while preserving each query's semantics.

This paper makes the following contributions.

\begin{enumerate}

\item
We introduce state-centric execution for dynamic concurrent workloads.
It treats operator state as shared state and makes that state the unit of sharing.

\item
We design the runtime mechanisms that support dynamic folding over
shared state.
Per-query state lenses use coverage metadata and per-query visibility metadata to
define what each query may observe, while query grafting attaches
arriving queries to compatible shared state.

\item We implement \SYSNAME{} and evaluate it on dynamic concurrent
workloads generated from TPC-H templates.
\SYSNAME{} achieves up to 2.17 times higher throughput than the isolated baseline
and, under overloaded open-loop arrivals, reduces P95 response time to as low as
0.17 times the isolated baseline's P95 response time.

\end{enumerate}

The remainder of the paper is organized as follows.
Section~\ref{sec:background-motivation} discusses prior ways to share analytical
work and motivates shared state as the unit of sharing.
Section~\ref{sec:overview} introduces GraftDB's execution model and
illustrates dynamic folding over shared state.
Section~\ref{sec:dynamic-sharing} defines per-query state lenses over shared state.
Section~\ref{sec:runtime-management} describes query grafting and the runtime scheduling
of folded queries.
Section~\ref{sec:evaluation} evaluates GraftDB on dynamic concurrent
workloads and separates the effects of its dynamic-folding mechanisms.
Section~\ref{sec:related-work} discusses related work, and
Section~\ref{sec:conclusion} concludes.

\section{Background and Motivation}
\label{sec:background-motivation}

Prior analytical work sharing reuses work through stored results,
known query sets, and scans, pipelines, or joins coordinated during
execution.
Dynamic concurrent workloads further expose partially produced
operator state as a sharing unit.

Existing mechanisms reduce redundant analytical work by choosing a unit that can be
shared.
Stored-result and materialized-view approaches share saved artifacts once compatible
work has been completed and recorded
\cite{conf/pods/LevyMSS95,conf/sigmod/GoldsteinL01,conf/sigmod/IvanovaKNG09,
conf/icde/NagelBV13}.
Multi-query optimization and batched shared-execution approaches share
plans, subplans, or execution cycles chosen for a known query set before
that shared unit runs
\cite{conf/sigmod/Finkelstein82,journals/tods/Sellis88,conf/sigmod/RoySSB00,
journals/pvldb/GiannikisAK12,journals/pvldb/GiannikisMAK14,
journals/pvldb/MakreshanskiGAK16}.
Runtime-sharing systems move the sharing decision into execution by coordinating
scans, operator pipelines, shared joins, or data-routing paths while queries are running
\cite{conf/sigmod/HarizopoulosSA05,conf/vldb/ZukowskiHNB07,
journals/pvldb/CandeaPV09,journals/pvldb/UnterbrunnerGAFK09,
conf/sigmod/ArumugamDJPP10}.
Together, these mechanisms show that the sharing unit matters:
saved artifacts, planned query sets, and running execution structures expose different
places where redundant analytical work can be reduced.

\begin{figure*}[t!]
  \centering
  \includegraphics[width=\textwidth]{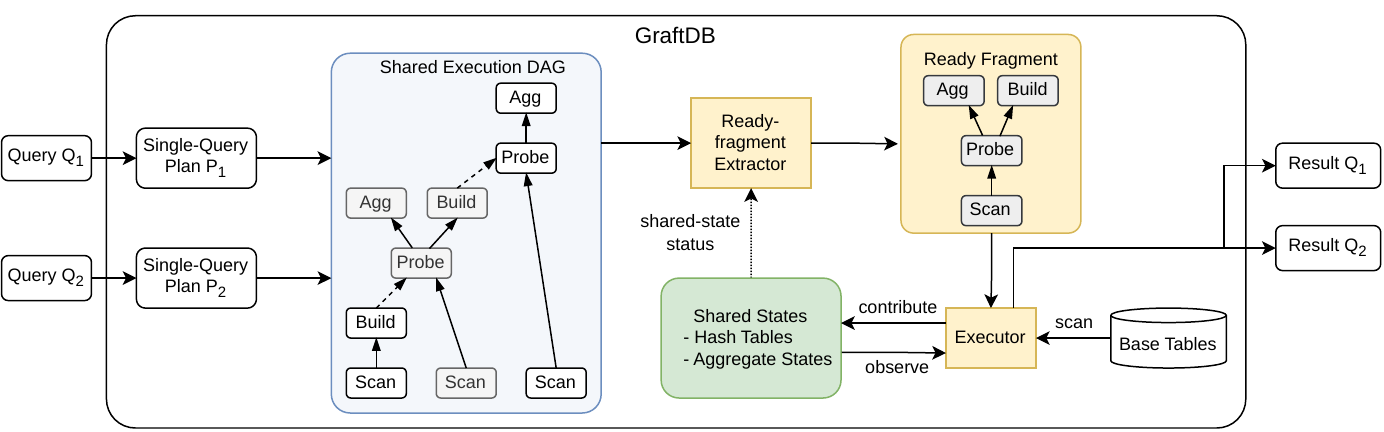}
  \Description{System overview showing online query arrivals becoming ordinary single-query
  plans, dynamic folding through attachment into a shared execution DAG,
  ready-fragment extraction using
  shared-state status and source progress, executor scans over base tables, executor observes
  and contributes to shared state, and per-query results.}
  \caption{State-centric execution architecture.
  \SYSNAME{} first represents arriving queries as ordinary single-query plans and
  attaches them to a shared execution DAG.
  \SYSNAME{} extracts ready fragments using the current status of shared states and input
  sources, then executes those fragments by scanning base tables, contributing to shared hash
  tables and aggregate states, and observing shared state through per-query state lenses.}
  \label{fig:section3-system-overview}
\end{figure*}

Stateful operators expose another unit of reuse inside a running
execution. A hash table records build-side work that later probes
need, and aggregate state records groups and results that
later aggregate computations need.
When concurrent analytical queries require overlapping state, state built for one query may also
be useful to a later query.
If that state remains private to the query that first created it, the later query must rebuild
overlapping work.

This state differs from both a stored result and a single logical
result. It may still be incomplete and changing when a later query
arrives, so the query must know what is already present and what
the running execution can still add. It may also have different
meanings for different queries because each query may need a different
logical slice of the same physical state.
These properties motivate state-centric execution, where sharing is organized
around shared state already produced by a running execution and
the work that can still contribute to it.

\section{\SYSNAME{} Overview: State-centric Execution for Dynamic Folding}
\label{sec:overview}

\subsection{State-centric Execution Architecture}
\label{sec:shared-state-model}

State-centric execution is \SYSNAME{}'s execution model for dynamic
folding.
It organizes a dynamic concurrent workload as one evolving shared
execution over shared state.
Stateful operators maintain hash tables and aggregate states as shared
state that multiple queries can observe or contribute to.
This organization lets a later-arriving query attach to the running
execution by observing compatible state already built there and sharing
producer work that can still contribute to it.

Figure~\ref{fig:section3-system-overview} shows the execution
architecture of state-centric execution.
First, \SYSNAME{} represents arriving queries as ordinary single-query plans.
It then attaches those plans to the shared execution DAG, which records
the operator work that remains in the shared execution.
The DAG contains ordinary operator work, work that can contribute to shared
states, and state-lens observations that depend on those states.
As the shared execution advances, the ready-fragment extractor uses
the DAG and the current status of shared states and input sources to
select fragments whose requirements are satisfied.
The executor advances those fragments by scanning base tables,
contributing to shared hash tables and aggregate states, and applying per-query
state lenses to shared state to produce each query's result.

A query does not observe the whole physical contents of a shared state.
Instead, a per-query state lens selects the entries or state fragments visible under that query's
semantics, without materializing a separate per-query copy.
Different queries can therefore share the same physical state while receiving
only the state fragments visible through their own state lenses.

\subsection{Query Scope and Sharing Units}

\SYSNAME{} targets finite analytical \texttt{SELECT} queries that can be represented as
acyclic relational operator plans built from base-table scans, selections, projections,
hash joins, and aggregations.
In this paper, a query is eligible to fold into a shared execution only when it reads
the same read-only database snapshot as that execution.
Coverage metadata therefore describes completeness over one stable input version.
Queries may differ in predicates, constants, and surrounding operator structure within
this plan class.
Dynamic folding uses compatible stateful boundaries as the comparison unit, so different
query plans can still share state where their stateful requirements are compatible.

The current design shares two kinds of operator state:
hash-build state and aggregate state.
For hash joins, a later query may observe represented build-side extents through a
per-query state lens, while admitted producer work contributes residual build-side extents
to the same hash-build state.
Probe-side input remains a consumer-side data-flow requirement and does not itself define
residual build-side work.

For aggregations, sharing is admitted only under exact aggregate identity:
the aggregate input, per-query input condition, grouping keys, aggregate functions, and
distinct-argument semantics must match.
When this identity holds, compatible queries may share aggregate production or observe a
completed aggregate state.
A completed aggregate state remains tied to that identity; different predicates or
grouping rules produce separate aggregate states.

\subsection{Folding a Later Query into Shared State}
\label{sec:hash-join-attachment-example}

\begin{figure}[!t]
  \centering
  \includegraphics[width=\linewidth]{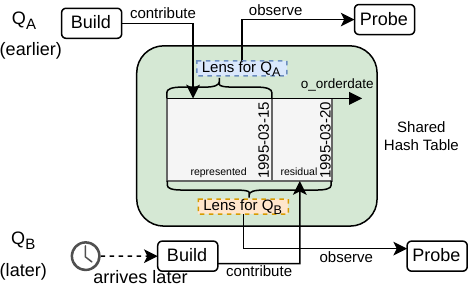}
  \Description{Diagram of two TPC-H Q3-derived queries sharing one order-side hash table.
  Query A arrives earlier and constructs the initial date range.
  Query B arrives later, adds rows from the missing date band to the same hash table,
  and observes the table through its own state lens.
  Query A and Query B observe the shared hash table through state lenses defined by
  their date predicates.}
  \caption{Folding a later query into a shared order-side hash table.
  \query{A} arrives first and constructs the initial order-date range.
  When \query{B} arrives later with a broader order-date predicate, \SYSNAME{} lets
  \query{B} observe the already-built range as its represented extent and contributes the
  missing date band through residual producer work to the same hash table.
  Each query receives only the entries selected by its state lens.}
  \label{fig:section3-shared-state-attachment}
\end{figure}

Dynamic folding becomes concrete when an arriving query can observe state maintained by
the shared execution.
At each stateful boundary, \SYSNAME{} compares the query's required state-side input
with a compatible shared state and partitions that input into three extents.
The represented extent is already complete in the selected state; the residual extent
can still be contributed to that state by an admitted producer path; and the unattached
extent remains ordinary-plan work.
This partition lets a later query observe state built before it arrived and share
producer work performed after it arrived.

For a hash join, the attachment decision concerns the build-side rows
represented by the hash table.
The probe rows have a different role.
They do not contribute build entries to the hash table; they drive lookups into it.
A folded hash-probe step can therefore run only after the shared hash
table represents the represented and residual build-side extents assigned
to the query, and after the probe rows for that query are available.

We use two TPC-H Q3-derived queries as a running hash-join instance.
Figure~\ref{fig:section3-shared-state-attachment} focuses on the
order-side hash table and the per-query state lenses over it.
The instance isolates one hash-join boundary so that already-present
state-side input, residual producer work, and probe rows are visible in one
place.
Both queries use the following parameterized SQL skeleton:

\begin{lstlisting}[style=sqlblock]
SELECT l_orderkey,
       sum(l_extendedprice * (1 - l_discount)) AS revenue,
       o_orderdate, o_shippriority
FROM customer, orders, lineitem
WHERE c_mktsegment = :segment
  AND c_custkey = o_custkey
  AND o_orderdate < :date
  AND l_orderkey = o_orderkey
  AND l_shipdate > :date
GROUP BY o_orderdate, o_shippriority, l_orderkey
\end{lstlisting}

Both queries use \texttt{:segment = 'BUILDING'}. Query $Q_A$
arrives first with \texttt{:date = DATE '1995-03-15'}, and query
$Q_B$ arrives later with \texttt{:date = DATE '1995-03-20'}.
In this example, $Q_B$ therefore has a broader order-side build predicate than
$Q_A$ but a narrower lineitem probe predicate.
For exposition, assume a fixed hash-join plan. The plan builds
customer-side state for customers satisfying
\texttt{c\_mktsegment} $=$ \texttt{'BUILDING'}, builds order-side
state for orders that match the customer state and satisfy
\texttt{o\_orderdate} $<$ \texttt{:date}, and then probes the
order-side hash table with lineitem rows satisfying
\texttt{l\_shipdate} $>$ \texttt{:date}.

On the order build side, $Q_B$ has the broader date predicate.
Every order row satisfying $Q_A$'s bound,
\texttt{o\_orderdate} $<$ \texttt{DATE '1995-03-15'}, also
satisfies $Q_B$'s bound,
\texttt{o\_orderdate} $<$ \texttt{DATE '1995-03-20'}. The range
already constructed for $Q_A$ is therefore part of $Q_B$'s represented
extent and is visible through $Q_B$'s state lens.

The remaining order-date band for $Q_B$, from
\texttt{DATE '1995-03-15'} to before
\texttt{DATE '1995-03-20'}, belongs to $Q_B$'s residual extent if the
current shared execution can still produce those order rows and add them
to the same order-side hash table.
Rows at or after \texttt{DATE '1995-03-20'} are outside $Q_B$'s
build requirement.
They are outside both $Q_B$'s represented and residual extents.
$Q_B$'s state lens therefore excludes them even if another
query later adds such rows to the same hash table.

The lineitem side is different because it drives lookups into the
order-side hash table rather than contributing entries to that table.
Since $Q_B$ uses the later ship-date bound, only lineitem rows with
\texttt{l\_shipdate} $>$ \texttt{DATE '1995-03-20'} drive
$Q_B$'s hash-probe step.
Rows with ship dates after \texttt{DATE '1995-03-15'} and at or
before \texttt{DATE '1995-03-20'} can drive $Q_A$'s hash-probe step, but they
are not part of $Q_B$'s residual extent because they are not missing
order-side build rows.
Shared scans and filters tag rows with the queries whose predicates
they satisfy.
$Q_B$'s shared hash-probe step becomes ready when the order-side hash
table contains the build rows assigned to $Q_B$ and the lineitem rows
for $Q_B$ are available.

\section{State-Lens Observation over Shared State}
\label{sec:dynamic-sharing}

A state-lens observation is the point at which one query observes shared physical
state under that query's semantics.
The lens combines two checks.
Coverage metadata identifies when the selected state represents the relevant input,
including the absence information needed for valid no-match results.
Per-query visibility metadata identifies which materialized rows or state fragments may
be emitted to the query.

\subsection{Lens Descriptors and Extent Assignment}
\label{sec:lens-descriptors}

At a stateful boundary \(b\) of query \(q\), \SYSNAME{} considers a candidate
shared state \(S\).
The candidate check uses three objects:
a lens descriptor that describes the operator state needed by \(q\), the state-side
input that \(S\) must represent, and any consumer-side input needed to drive the
state-consuming operator.

For a hash-probe step, the lens descriptor names the build relation, build keys,
build-side predicates, payload layout, and required upstream state.
For aggregate state, the lens descriptor records the aggregate identity used to decide
whether the state can support the query.

The state-side input is the set of input occurrences that \(S\) must represent for \(q\).
\SYSNAME{} identifies an occurrence by its derivation, not only by its payload value.
Two equal-valued tuples therefore remain distinct when they come from different base rows
or join derivations.
This keeps duplicate-sensitive row identity explicit when multiple per-query state-lens
observations share one state.

After selecting \(S\), \SYSNAME{} partitions the attached state-side input into
three disjoint extents.
The represented extent \(E_{\mathit{rep}}\) is the portion already represented by
\(S\) and assigned for observation through the state lens.
The residual extent \(E_{\mathit{res}}\) is not yet represented, but an admitted producer
path can produce those occurrences into \(S\) before \(q\) observes the state.
The unattached extent \(E_{\mathit{un}}\) is not assigned to that lens and is executed
as ordinary-plan work.
The state-lens observation through \(S\) is defined over
\(E_{\mathit{rep}} \cup E_{\mathit{res}}\).

This assignment applies only to state-side input.
Hash-probe steps also require probe-side input.
The hash table represents build-side input, while probe rows drive lookups into that
table.
A hash table may therefore be ready for a query's build-side requirement before the
corresponding probe-side input is available.

\subsection{Predicate and Visibility Checks}
\label{sec:predicate-visibility-checks}

Predicate compatibility is treated as containment rather than syntactic
equality.
The prototype stores state-side predicates in lens descriptors and the predicate
component of coverage metadata as normalized predicate ASTs.
For predicates \(P\) and \(Q\) over comparable state-side attributes, \SYSNAME{}
writes \(\mathsf{Prove}(P \Rightarrow Q)\) for a conservative proof that every
occurrence satisfying \(P\) also satisfies \(Q\).
For hash-build state, let \(P_R\) describe the predicate of a candidate represented extent
\(R\), let \(B_q\) be query \(q\)'s required build-side predicate, and let \(C_S\)
describe an extent that state \(S\)'s coverage metadata records as complete.
Assigning \(R\) to the represented extent requires proving both
\(\mathsf{Prove}(P_R \Rightarrow B_q)\) and
\(\mathsf{Prove}(P_R \Rightarrow C_S)\).

The checker is sound but incomplete.
It implements \(\mathsf{Prove}(P \Rightarrow Q)\) by simplifying
\(\neg P \vee Q\) within a supported deterministic predicate fragment.
The implemented cases cover conjunctions of deterministic comparisons between retained
attributes and constants, together with limited Boolean simplifications introduced by the
implication check.
The checker canonicalizes equality predicates and lower and upper bounds on each
retained attribute, then applies per-attribute range-containment rules independently
over comparable scalar domains.
Constant arithmetic already normalized into the predicate AST is handled as part of the
same check.
Predicate forms outside this fragment, including unsupported NULL-sensitive predicate
forms, are treated as unproven.
Unproven obligations are not used to classify an extent as represented:
the corresponding input is considered for residual production, or left as ordinary-plan work
if no admitted producer path can supply it.
Thus failing to prove a valid implication may reduce sharing, but it cannot make an
unsafe state-lens observation admissible.

A visibility check also requires the relevant lens predicates to be evaluable on the
retained state.
Each predicate AST records the attributes it references, written
\(\mathsf{FV}(P)\).
A lens predicate \(P\) is evaluable on entries of state \(S\) only when
\(\mathsf{FV}(P) \subseteq \mathsf{RetainedAttrs}(S)\), where retained attributes
include attributes stored with state entries and derivation metadata.
If a later query requires an additional narrower predicate whose referenced attributes
are not retained, \SYSNAME{} does not classify that part of the state-side extent as
already represented.
It belongs to the residual extent if an admitted producer path can supply it, or to the
unattached extent otherwise.

Rows and state entries carry per-query visibility metadata.
Filters update that metadata according to each query's predicate, and projections preserve
derivation identity and visibility.
A per-query state lens is therefore a visibility boundary rather than a separate physical
copy of state.
A physical producer or consumer step may serve several queries, provided each emitted row
or state entry passes the corresponding visibility check.

\subsection{Hash Tables as Shared Build State}
\label{sec:hash-build-state}

A shared hash-build state contains a hash-table signature, coverage metadata, and hash
entries.
The hash-table signature fixes the build relation, build keys, payload layout, and required
upstream state.
Coverage metadata records the build-side extent for which the table is complete, while
hash entries store keys, payload values, derivation identifiers, and entry-level
visibility metadata.

\begin{figure}[t]
  \centering
  \includegraphics[width=\linewidth]{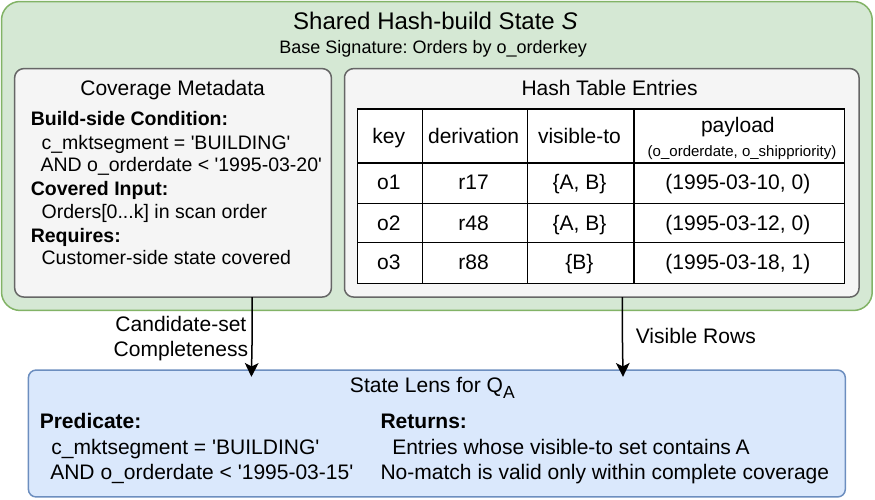}
  \caption{Shared hash-build state separates coverage from entry visibility.
  Coverage metadata describes the build-side predicate, processed input range, and upstream
  requirement that make a candidate set complete.
  Hash entries store concrete materialized rows together with per-query visibility metadata.
  A per-query state lens uses coverage metadata to determine where complete match sets and
  no-match results are meaningful, and visibility metadata to select the rows visible
  to that query.}
  \Description{A schematic of shared hash-build state with a coverage-metadata panel, a hash table
  entries panel, and a Query A state lens.
  The coverage-metadata panel shows a broader build-side condition, a processed order-input range,
  and a required customer-side dependency.
  The entry panel records keys, derivations, per-query visibility metadata, and payload values.
  Arrows show that candidate-set completeness comes from coverage metadata, while visible rows come
  from hash table entries.}
  \label{fig:hash-build-state-coverage}
\end{figure}

Figure~\ref{fig:hash-build-state-coverage} shows this separation on the order-side
table.
The table's coverage metadata describes the represented order input, the build-side
condition, and the required customer-side state.
It does not enumerate returned rows.
Instead, it records where the build-side extent is complete for compatible queries.

Hash entries provide positive matches, while coverage metadata makes absence meaningful.
For a probe occurrence visible to query \(q\), the hash-probe step computes the probe key,
finds candidate entries, and keeps only entries visible through \(q\)'s state lens.
A visible entry can justify an output row because the matching entry is present and
visible.
If no visible matching entry exists, the hash-probe step may report no match only when
coverage metadata records the relevant build-side extent as complete for \(q\).
When a probe row is visible to several queries and the lookup target is shared, one
physical hash-probe step can test candidate entries once and route each matching entry to
the queries for which the visibility check succeeds.

The same visibility discipline applies on the build path.
When a produced build row is visible to several queries and targets the same hash table,
\SYSNAME{} stores one build entry and records the visibility needed by those queries.
The visibility sets in Figure~\ref{fig:hash-build-state-coverage} are logical notation:
the runtime need not maintain a mutable set in every hash entry or rewrite existing entries
when a later query is admitted.
If a later query observes an already represented extent, \SYSNAME{} may record visibility
at the state level for that extent.
This state-level visibility is extent-scoped: it does not make every entry in the hash
table visible to the query.
Residual build-side extents for several admitted queries can therefore pass through the same
physical producer path when the path and target table are shared, while each query later
receives only the entries visible through its state lens.

Before opening a hash-probe step, \SYSNAME{} first checks exact non-predicate
compatibility:
the relation, keys, payload layout, and required upstream state must match the candidate
table.
Predicate containment and evaluability are checked as described in
Section~\ref{sec:predicate-visibility-checks}.
When the selected table represents the assigned build-side extent and the probe-side input
is available, the state-lens observation runs as a per-query hash-probe step and emits
joined rows allowed by the query's predicates and join rule.

\subsection{Residual Production through Shared Scans}

Residual production contributes to a selected state through an admitted state-producing
path.
The path must be compatible with the selected state's descriptor and operator rule.
The residual extent for a boundary is the state-side portion not yet represented by
the selected state that an admitted path can contribute to that state.
The path determines the granularity at which the residual extent can be contributed
to the selected state.

Shared scans determine which admitted paths can still receive base input.
\SYSNAME{}'s shared scans run in cycles over their input.
Depending on the current scan state, the runtime may register a path that receives rows
remaining in the current cycle, or a path that waits for a later cycle to deliver rows to
the selected state.
Rows before the current scan position do not belong to the residual extent merely because
the scan will revisit input.
They belong to the residual extent only when the attachment records a path that will
deliver them to the selected state.

An admitted scan path can contribute residual input for several queries when their paths
target the same selected state.
As the scan produces a row, filters update per-query visibility, and the shared execution
routes the row once through the common producer path while preserving the visibility needed
by each state lens.

Scan progress and coverage metadata answer different questions.
Scan progress determines when base rows can next be delivered by the shared scan, whereas
coverage metadata determines which state-side extent the selected state already represents
and where complete match sets or no-match results are meaningful.
A state-lens observation opens only after the selected state represents the observation's
represented and residual extents.

\subsection{Aggregate Sharing}

Aggregation follows the same rule with a stricter identity requirement.
A query may observe completed aggregate state only when the aggregate input, per-query
input condition, grouping keys, aggregate functions, and distinct-argument semantics
match.
Under this exact aggregate identity, the query observes a completed aggregate state with
the identity recorded in the lens descriptor.

Active queries can also share aggregate production when the current shared execution
admits live sharing under the same aggregate identity.
In that case, all compatible queries for that identity share one aggregate producer and
one aggregate state.
Aggregate sharing is exact-match active sharing:
the aggregate descriptor fixes this identity, and compatible aggregate work contributes
to the same state.
Different aggregate definitions remain separate aggregate states.
Unlike hash-build state, aggregate state collapses input occurrences into group
accumulators, so a completed aggregate cannot be repartitioned under a different predicate
or grouping rule without additional provenance or partial-aggregate state.
This boundary preserves exact aggregate identity while still allowing compatible aggregate
work to share one state.

\subsection{Opening State-Lens Observations}

A state-lens observation is admitted for a query when the state-side occurrences for the
selected state have a single assignment and a defined per-query state lens.
The represented extent contains occurrences already represented by the selected state.
The residual extent contains admitted occurrences that are produced into the same state
before the observation opens.

This condition keeps duplicate-sensitive row identity explicit.
Occurrences identified by derivation remain distinct, and \SYSNAME{} does not duplicate
them across represented and residual extents.
It also restricts no-match results because absence is used only within the complete extent
described by coverage metadata.
If descriptor compatibility, extent completeness, aggregate identity, or an admitted path for a
missing extent cannot be established, that extent is not assigned to the selected state
lens.

The admission condition is stated per query, but it does not require per-query physical
state or per-query physical execution.
The shared execution may satisfy several per-query admission conditions through one scan
step, one admitted state-producing step, or one probe step, as long as visibility metadata
routes each row and state entry to its allowed queries.

A state-lens observation becomes state-ready when the selected state covers that lens's
represented and residual extents.
The state-consuming operator is scheduled only after ready-fragment extraction finds a
data-flow path that supplies the required consumer-side input.
It then emits rows for the assigned state-side input through that state lens.
The boundary combines those rows with outputs produced by ordinary-plan work for
the unattached extent.

\section{Query Grafting and Scheduling}
\label{sec:runtime-management}

Query grafting installs DAG work for an admitted boundary.
When a query arrives, \SYSNAME{} compares each stateful boundary in its single-query DAG
with shared state maintained by the running execution.
For an admitted boundary, the runtime turns the boundary's state-side extent into three
assignments: a state-ref edge over the represented and residual extents, producer edges
that contribute the residual extent to the selected state, and ordinary-plan work for the
unattached extent.
The same DAG update installs a state-readiness gate on the state-ref edge.
This gate records when the selected state is ready for that edge; the consumer-side input
that drives the operator remains part of the DAG's data flow, and ready-fragment
extraction handles that input.

\subsection{The Shared Execution DAG}
\label{sec:runtime-shared-dag}

The shared execution DAG records the work that can still advance in the running execution.
A node represents an operator instance together with the queries assigned to that instance.
A data edge carries rows between operators.
A state-ref edge connects a state-consuming operator to shared state through a
query-specific state-readiness gate.
For a DAG \(G\), we write \(\mathsf{DataEdge}(G)\) for its data edges,
\(\mathsf{StateRefEdge}(G)\) for its state-ref edges, and \(\mathsf{DepEdge}(G)\) for
their union.

\begin{figure}[t]
  \centering
  \includegraphics[width=\linewidth]{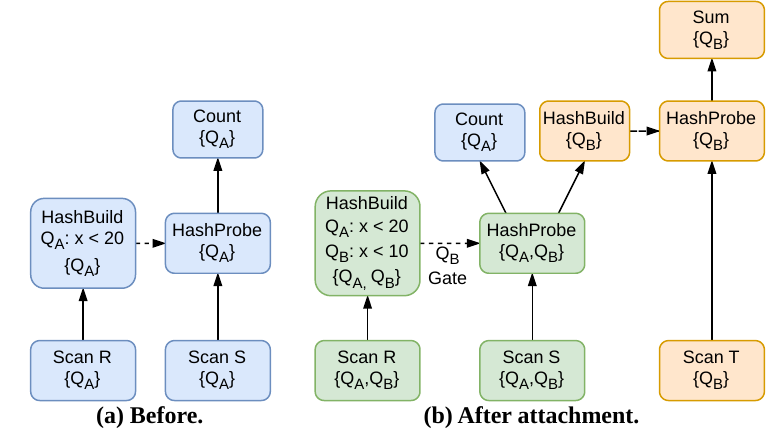}
  \Description{Shared execution DAG transition for two non-identical queries over relations R, S,
  and T. Before attachment, Query A has a scan, hash-build, hash-probe, and count path.
  After attachment, Query B shares the R and S scans, the hash-build state, and the prefix up to
  the first hash-probe, while its T-side suffix remains separate.
  The shared prefix records different predicates for Query A and Query B, and Query B's state-ref
  edge is labeled with a Query B gate.}
  \caption{Shared execution DAG update for two non-identical queries.
  Solid edges carry row flow, dashed edges carry state-ref dependencies, and
  the \query{B} gate labels the state-readiness condition on \query{B}'s state-ref edge.}
  \label{fig:dag-transition}
\end{figure}

The DAG also records the attachment decisions made for arriving queries: residual producer
edges, ordinary-plan assignments, and state-ref edges guarded by per-query state lenses.
Figure~\ref{fig:dag-transition} illustrates one DAG update.
When \query{B} arrives with a narrower \(x < 10\) requirement and a \(T\)-side suffix,
\SYSNAME{} attaches its compatible prefix to the existing hash-build state, records a
\query{B}-specific state-readiness condition on the state-ref edge, and leaves the
\query{B}-only suffix as separate operator work.
The gate label denotes readiness of the selected shared state, not a physical operator and
not the availability of the probe-side input.

\subsection{Query-grafting Admission}
\label{sec:runtime-dynamic-attachment}

Admission is local to a stateful boundary.
\SYSNAME{} first compiles the arriving query \(q\) into an ordinary single-query DAG.
For each stateful boundary \(b\), it forms a lens descriptor \(d=(a,\rho)\), where the
lens signature \(a\) records state-side identity and lens conditions:
relations, keys, predicates, grouping attributes, aggregate identity, and required
upstream state.
The operator rule \(\rho\) records which matches, groups, or outputs \(q\) can
derive from a compatible state.
A candidate shared state \(S\) can be selected only when it provides a compatible
signature and the metadata needed to form \(q\)'s per-query state lens.

For each stateful boundary, the prototype derives a canonical state signature and uses
the shared execution's signature index to select a corresponding live or retained shared
state when the state-ref reuse checks pass.
Algorithm~\ref{alg:admit-boundary} then applies coverage and visibility checks to
assign the boundary's state-side input to represented, residual, and unattached extents.

For an admissible candidate, \SYSNAME{} partitions the state-side extent required by \(b\).
The represented extent \(R_{\mathit{rep}}\) is already represented by \(S\).
For hash-build state, this means that the table's coverage metadata describes the
corresponding build-side extent as complete.
For aggregate state, this means that the aggregate state has the exact aggregate identity
and the physical aggregate state binding used by the shared DAG.
Missing occurrences that an admitted producer path can still deliver to \(S\) become the
residual extent \(R_{\mathit{res}}\); the rest becomes the unattached extent
\(R_{\mathit{un}}\).
The partition is over derivation-identified occurrences, so equal payload tuples are not
merged by the assignment.

The partition has direct operational effects in the DAG: \SYSNAME{} installs a state-ref
edge for \(R_{\mathit{rep}} \cup R_{\mathit{res}}\), installs residual producer edges
targeting \(S\), and records \(R_{\mathit{un}}\) as ordinary-plan work before the
state-consuming work is scheduled.
Because the assignment is made per boundary, one query can observe shared state at one
operator, contribute a residual extent for another, and keep unrelated work as
ordinary-plan work for that query.

Algorithm~\ref{alg:admit-boundary} shows the query-grafting admission decision for one
boundary and one candidate state.
The function either rejects the candidate, leaves the boundary as ordinary-plan work, or
installs a state-ref edge whose state dependency is open immediately or gated.
Consumer-side data availability is represented by data edges and checked during
ready-fragment extraction.

\begin{algorithm}[t]
\caption{Query-grafting admission for one boundary and one candidate state.}
\Description{Pseudocode showing how query-grafting admission checks a candidate shared state,
partitions a boundary extent, installs residual producers and a state-ref edge, and returns an
immediate or gated state-ref edge.}
\label{alg:admit-boundary}
\begin{algorithmic}[1]
\Function{AdmitBoundary}{$q,b,S,G,t$}
    \State $v \gets \Call{CheckLensCompatibility}{q,b,S,G}$
        \label{line:admit-check}
    \If{$v = \bot$}
        \label{line:admit-invalid-if}
        \State \Return \Call{NoAttachment}{$S$}
            \label{line:admit-noattach}
    \EndIf

    \State $X \gets \Call{PartitionStateExtent}{q,b,S,G,v}$
        \label{line:admit-partition}
    \State $(R_{\mathit{rep}},R_{\mathit{res}},R_{\mathit{un}},
        \mathcal{P}_{\mathrm{res}})
        \gets X$
    \State $\Call{AssignOrdinarySource}{q,b,R_{\mathit{un}},G}$
        \label{line:admit-assign-un}

    \If{$R_{\mathit{rep}} \cup R_{\mathit{res}} = \emptyset$}
        \label{line:admit-empty-if}
        \State \Return \Call{OrdinaryOnly}{$q,b$}
            \label{line:admit-empty-return}
    \EndIf

    \If{$R_{\mathit{res}} \neq \emptyset$}
        \label{line:admit-res-if}
        \State $\Call{InstallResidualProducers}
            {q,b,S,R_{\mathit{res}},\mathcal{P}_{\mathrm{res}},G}$
            \label{line:admit-install-res}
    \EndIf

    \State $r \gets (q,b,v)$
        \label{line:admit-state-ref-record}
    \State $O \gets
        \Call{InstallStateRef}{r,S,R_{\mathit{rep}} \cup R_{\mathit{res}},G}$
        \label{line:admit-install-state-ref}

    \If{\Call{Open}{$O,t$}}
        \label{line:admit-open-if}
        \State \Return \Call{OpenStateRef}{$r,O$}
            \label{line:admit-open-return}
    \Else
        \State \Return \Call{GatedStateRef}{$r,O$}
            \label{line:admit-gated-return}
    \EndIf
\EndFunction

\Function{PartitionStateExtent}{$q,b,S,G,v$}
    \State $R \gets \Call{StateExtent}{q,b}$
        \label{line:partition-stateextent}
    \State $R_{\mathit{rep}} \gets R \cap \Call{RepresentedExtent}{S,v,G}$
        \label{line:partition-rep}
    \State $R_{\mathit{miss}} \gets R \setminus R_{\mathit{rep}}$
        \label{line:partition-miss}
    \State $\mathcal{P}_{\mathrm{res}} \gets
        \Call{AdmissibleProducerPaths}{R_{\mathit{miss}},S,G,v}$
        \label{line:partition-paths}
    \State $R_{\mathit{res}} \gets
        R_{\mathit{miss}} \cap \Call{Extent}{\mathcal{P}_{\mathrm{res}}}$
        \label{line:partition-res}
    \State $R_{\mathit{un}} \gets R_{\mathit{miss}} \setminus R_{\mathit{res}}$
        \label{line:partition-un}
    \State \Return $(R_{\mathit{rep}},R_{\mathit{res}},R_{\mathit{un}},
        \mathcal{P}_{\mathrm{res}})$
        \label{line:partition-return}
\EndFunction
\end{algorithmic}
\end{algorithm}

The first two blocks make the admission test operational.
Line~\ref{line:admit-check} checks the exact non-predicate conditions needed for
\(S\) to support \(q\)'s state lens at boundary \(b\), including state kind,
relation, key and payload layout, aggregate identity when applicable, and required
upstream state.
It also records the predicate-containment and evaluability obligations for that lens.
If this check fails, the rejection is local to this candidate state
(lines~\ref{line:admit-invalid-if}--\ref{line:admit-noattach}).

The partition step then evaluates those obligations using the predicate-containment and
evaluability checks from Section~\ref{sec:predicate-visibility-checks}.
Line~\ref{line:partition-rep} places only extents whose obligations are proven in
\(R_{\mathit{rep}}\).
A missing extent is placed in \(R_{\mathit{res}}\) only when
\textsc{AdmissibleProducerPaths} finds a path that can still deliver it into \(S\)
(lines~\ref{line:partition-paths}--\ref{line:partition-res}).
All other occurrences are placed in \(R_{\mathit{un}}\)
(line~\ref{line:partition-un}).
\textsc{AdmitBoundary} records the unattached extent as ordinary-plan work before
installing any state-ref edge (line~\ref{line:admit-assign-un}).
Thus Algorithm~\ref{alg:admit-boundary} treats unsupported predicate reasoning as lost
sharing rather than unsafe sharing.

The final block installs the runtime obligations.
If no represented or residual extent is assigned to the selected state, the boundary remains
ordinary-only (lines~\ref{line:admit-empty-if}--\ref{line:admit-empty-return}).
Otherwise, \SYSNAME{} installs producer obligations for a nonempty residual extent into
\(S\) (lines~\ref{line:admit-res-if}--\ref{line:admit-install-res}), and the admitted
boundary installs a state-ref edge over \(R_{\mathit{rep}} \cup R_{\mathit{res}}\)
(lines~\ref{line:admit-state-ref-record}--\ref{line:admit-install-state-ref}).
The final branch reports whether that state dependency is already open or still gated
(lines~\ref{line:admit-open-if}--\ref{line:admit-gated-return}).
Consumer-side input is not part of this admission gate; it is enforced later by data-edge
reachability during ready-fragment extraction.

\subsection{State-readiness Gates}

A state-readiness gate records the state dependency of an admitted state-ref edge.
For an admitted state-ref \(r=(q,b,v)\) through shared state \(S\), the gate \(O=(r,S,R)\)
records the selected state and the assigned state-side extent.
For a hash-probe step, \(R\) is the build-side extent that the hash table must represent.
The probe-side input remains on the DAG's data-flow edges.

At scheduling time \(t\), the gate is open when the selected shared state is ready for
the assigned state-side extent:
\[
  \mathsf{open}_t(O) \equiv \mathsf{stateReady}_t(S,r,R).
\]
For hash-build state, the selected state's coverage metadata describes the assigned
build-side extent as complete.
For aggregate state, readiness follows exact aggregate identity, the physical aggregate
state binding used by the shared DAG, and completion of the required aggregate state.
An open state-ref edge may still wait for data-flow reachability before it appears in a
ready fragment.

\subsection{Ready Fragment Scheduling}

Scheduling operates on active node-query pairs, not on whole nodes.
For a state-producing node, query \(q\) is active while producer work assigned to \(q\)
remains pending.
For a state-consuming node, \(q\) passes the state-dependency filter when every state-ref
gate for that node-query pair is open.
The node-query pair still appears in a ready fragment only if it remains on a ready
data-flow path after graph pruning.
Ordinary nodes keep the assigned queries recorded by their source assignments.

Algorithm~\ref{alg:extract-ready-fragments} extracts ready work from the current DAG at
scheduling time \(t\).
It first computes active node-query pairs using producer obligations and state-ref
gates, then prunes the resulting graph by data-edge reachability.

\begin{algorithm}[t]
\caption{Extracting ready fragments from active assignments.}
\Description{Pseudocode showing how the scheduler computes active node-query pairs, removes
blocked pairs, and groups the remaining DAG into ready fragments.}
\label{alg:extract-ready-fragments}
\begin{algorithmic}[1]
\Function{ExtractReadyFragments}{$G,t$}
    \State $P \gets \emptyset$
        \label{line:extract-init-p}
    \ForAll{$n \in \Call{Nodes}{G}$}
        \label{line:extract-node-loop}
        \ForAll{$q \in \Call{ActiveAtNode}{n,G,t}$}
            \label{line:extract-active-call}
            \State $P \gets P \cup \{(n,q)\}$
                \label{line:extract-add-pair}
        \EndFor
    \EndFor

    \State $G_P \gets \Call{RestrictToPairs}{G,P}$
        \label{line:extract-restrict}
    \State $G_R \gets \Call{PruneByDataReachability}{G_P,t}$
        \label{line:extract-prune}

    \State $\mathcal{F} \gets \emptyset$
        \label{line:extract-init-fragments}
    \ForAll{$C \in \Call{WeakComponents}{G_R,\mathsf{DepEdge}(G_R)}$}
        \label{line:extract-components}
        \State $F \gets \Call{TopologicalOrder}{C,\mathsf{DataEdge}(C)}$
            \label{line:extract-topological}
        \State $\mathcal{F} \gets \mathcal{F} \cup \{F\}$
            \label{line:extract-add-fragment}
    \EndFor
    \State \Return $\mathcal{F}$
        \label{line:extract-return}
\EndFunction

\Function{ActiveAtNode}{$n,G,t$}
    \State $A \gets \emptyset$
        \label{line:active-init}
    \ForAll{$q \in \Call{AssignedQueries}{n}$}
        \label{line:active-assigned-queries}
        \If{\Call{ProducerInactive}{$n,q,G,t$}}
            \label{line:active-producer-test}
            \State \textbf{continue}
                \label{line:active-producer-continue}
        \EndIf
        \If{\Call{StateConsumerBlocked}{$n,q,G,t$}}
            \label{line:active-consumer-test}
            \State \textbf{continue}
                \label{line:active-consumer-continue}
        \EndIf
        \State $A \gets A \cup \{q\}$
            \label{line:active-add}
    \EndFor
    \State \Return $A$
        \label{line:active-return}
\EndFunction

\Function{ProducerInactive}{$n,q,G,t$}
    \If{\textbf{not} \Call{ProducesState}{$n$}}
        \label{line:producer-not-state}
        \State \Return \textsc{False}
            \label{line:producer-false}
    \EndIf
    \State \Return \textbf{not} \Call{ProducerWorkPending}{n,q,G,t}
        \label{line:producer-pending}
\EndFunction

\Function{StateConsumerBlocked}{$n,q,G,t$}
    \If{\textbf{not} \Call{ConsumesState}{$n$}}
        \label{line:consumer-not-state}
        \State \Return \textsc{False}
            \label{line:consumer-false}
    \EndIf
    \ForAll{$O \in \Call{Refs}{n,q,G}$}
        \label{line:consumer-refs}
        \If{\textbf{not} \Call{Open}{O,t}}
            \label{line:consumer-open-test}
            \State \Return \textsc{True}
                \label{line:consumer-blocked}
        \EndIf
    \EndFor
    \State \Return \textsc{False}
        \label{line:consumer-open}
\EndFunction
\end{algorithmic}
\end{algorithm}

The extractor first constructs active node-query pairs
(lines~\ref{line:extract-init-p}--\ref{line:extract-add-pair}).
\textsc{ActiveAtNode} starts from the node's assigned queries.
A producer node-query pair is removed by the test at
line~\ref{line:active-producer-test}; the helper in
lines~\ref{line:producer-not-state}--\ref{line:producer-pending} keeps state producers
active only while producer work is pending.
A state-consuming node-query pair is removed by the test at
line~\ref{line:active-consumer-test}; the helper in
lines~\ref{line:consumer-not-state}--\ref{line:consumer-open} passes it only when all
state-ref gates entering that node-query pair are open.

The extractor then restricts the DAG to active node-query pairs
(line~\ref{line:extract-restrict}) and prunes the restricted graph by data-edge
reachability (line~\ref{line:extract-prune}).
This pruning step is where consumer-side input availability is enforced.
The remaining graph is grouped over data and state-ref dependencies
(line~\ref{line:extract-components}), ordered along data edges
(line~\ref{line:extract-topological}), and emitted as ready fragments
(lines~\ref{line:extract-add-fragment}--\ref{line:extract-return}).

Together, admission and ready-fragment extraction maintain one assignment for each
state-side occurrence: it belongs to the represented extent, is contributed to the
selected state through residual producer work, or belongs to the unattached extent executed
as ordinary-plan work.
The state-ref gate and data-flow pruning then ensure that state-ref edges open only
after the selected state is ready and the consumer-side input is available.
These conditions give the core correctness argument for the state-lens observations
described in Section~\ref{sec:dynamic-sharing}:
each derivation-identified state-side occurrence is accounted for exactly once,
state-lens observations open only after the assigned state extent is complete, and
per-query visibility checks filter state entries and emitted rows.\footnote{A full
formalization of this argument requires a separate treatment.}
When a query completes, \SYSNAME{} removes that query's assignments from the DAG.
Referenced shared state remains available; unreferenced state can be released according
to the runtime's retention policy.

\section{Experimental Evaluation}
\label{sec:evaluation}

We evaluate \SYSNAME{} on TPC-H-derived dynamic concurrent
workloads~\cite{misc/tpc/TPCH301}.
The experiments measure performance across concurrency, arrival,
workload-skew, and data-scale settings, and separate the throughput
effect of the main dynamic-folding mechanisms.

\subsection{Experimental Setup}
\label{sec:evaluation-setup}

We evaluate a Rust prototype of \SYSNAME{}.
A query instance is a parameterized TPC-H template with concrete parameter values.
Unless otherwise stated, workloads sample Q1 and Q3--Q10,\footnote{We omit
TPC-H Q2 because it requires a correlated subquery, which is outside the
prototype's current SQL support.}
from a Zipf distribution with parameter $\alpha = 1$.
Template parameters are sampled uniformly from large benchmark domains.
Exact duplicate query instances are therefore rare; overlap mainly comes from
related templates and compatible operator requirements.

The prototype is available at
\url{https://github.com/dbc-utokyoiis/GraftDB}.
Experiments use a server with two \mbox{Intel} Xeon Gold 6132 CPUs at 2.60\,GHz,
96\,GB RAM, and 29\,TB RAID-6 HDD storage.
All Rust variants use one~thread.
The evaluation therefore focuses on inter-query concurrency under a single-worker
execution model.
Extending the prototype to intra-query parallel execution remains future work.
In the evaluated prototype, the runtime releases operator state once no query in the
shared execution references it, so the reported gains reflect temporal overlap with
running shared executions.

The main comparisons are within the Rust prototype.
\emph{Isolated} is the same engine with sharing disabled.
\emph{QPipe-OSP} is a same-engine implementation of QPipe's on-demand
simultaneous pipelining policy~\cite{conf/sigmod/HarizopoulosSA05}.
It shares scans and in-flight operator instances under identical operator
profiles, including predicates and pre-filters, without \SYSNAME{}'s
coverage-based observation of already-built state.
PostgreSQL 16.13 provides an external reference point.
PostgreSQL is pinned to one CPU core, with JIT compilation and parallel query
execution disabled, and joins restricted to hash joins.
For each TPC-H query template, the prototype uses a fixed physical plan
whose join order and operator sequence match PostgreSQL's EXPLAIN plan under
this configuration; workload parameters change only predicates and constants.
All Rust variants start from this plan before any sharing decision is applied.
Throughput is completed queries per hour.

\subsection{Dynamic Folding on TPC-H Q3}
\label{sec:evaluation-q3}

\begin{figure}[t]
  \centering
  \includegraphics[width=\linewidth]{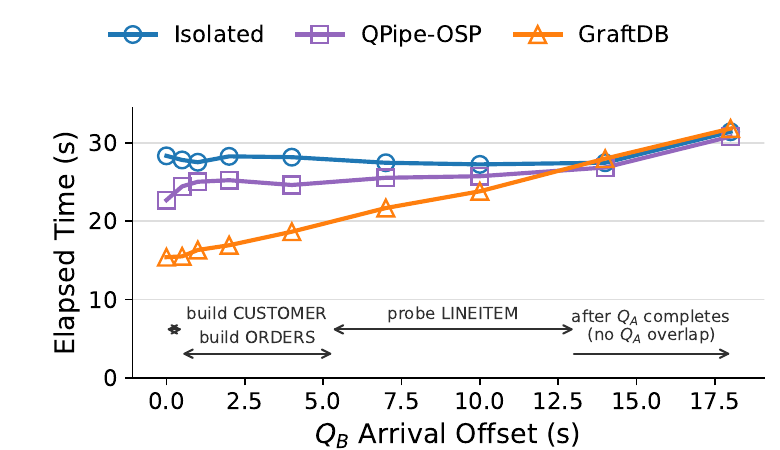}
  \caption{Elapsed time for two TPC-H Q3-derived queries as \query{B}'s arrival is delayed
  after \query{A}.
  Phase arrows show which part of \query{A} is running when \query{B} arrives.
  \SYSNAME{} shortens completion while \query{A}'s order-side state is live, then converges
  toward the baselines once \query{B} no longer overlaps with \query{A}.}
  \Description{A line chart with arrival offset on the x-axis and elapsed time on the y-axis.
  The chart compares Isolated, QPipe-OSP, and GraftDB, and phase arrows mark the customer build,
  orders build, lineitem probe, and no-overlap regions.}
  \label{fig:eval-q3-case-study}
\end{figure}

Figure~\ref{fig:eval-q3-case-study} uses the Q3-derived pair from
Section~\ref{sec:hash-join-attachment-example} at TPC-H SF10 and sweeps
\query{B}'s arrival offset after \query{A}.
The two queries use the same SQL skeleton and the same fixed parameters as in
Section~\ref{sec:hash-join-attachment-example}:
both use \texttt{:segment = 'BUILDING'}, \query{A} uses
\texttt{:date = DATE '1995-03-15'}, and \query{B} uses
\texttt{:date = DATE '1995-03-20'}.
The y-axis is elapsed time from the start of \query{A} until both queries
complete.

\emph{Isolated} executes independent physical plans.
Changing \query{B}'s arrival time changes temporal overlap between the two
executions, but it does not change the work assigned to either query.
While the two executions still overlap, the total elapsed time stays close to
one independent pair of Q3 executions.
After \query{A} finishes, the total elapsed time rises with the non-overlapped
suffix.

\emph{QPipe-OSP} shares scans and, for nearly simultaneous arrivals, the
customer build because both queries have the same customer-build profile,
including \texttt{c\_mktsegment = 'BUILDING'}.
After that phase, the order-side build and lineitem/probe profiles differ due
to the different date bounds, so \emph{QPipe-OSP} mainly benefits from scan
sharing.
Its elapsed time therefore stays below \emph{Isolated} but remains almost flat
over most arrival offsets.

\SYSNAME{} shares the same customer-build opportunity and also folds \query{B}
into the order-side state when that state can provide a represented extent for
\query{B}'s build-side requirement.
For this pair, \query{B}'s order predicate is broader than \query{A}'s.
The order hash table produced for \query{A} already represents the prefix of
\query{B}'s state-side extent shared by both queries, and \SYSNAME{} registers
the still-missing date band as the residual build-side extent when the current shared
execution can still contribute it to the order-side state.
The lineitem predicate moves in the opposite direction:
\query{B}'s later date narrows the probe-side input, so rows needed only by
\query{A} are not part of \query{B}'s residual extent.
At zero offset, elapsed time falls from 28.4\,s under \emph{Isolated} to
15.4\,s under \SYSNAME{}.
After \query{A} completes, \SYSNAME{} converges toward the same no-overlap range
as the baselines.

\begin{figure}[t]
  \centering
  \includegraphics[width=\linewidth]{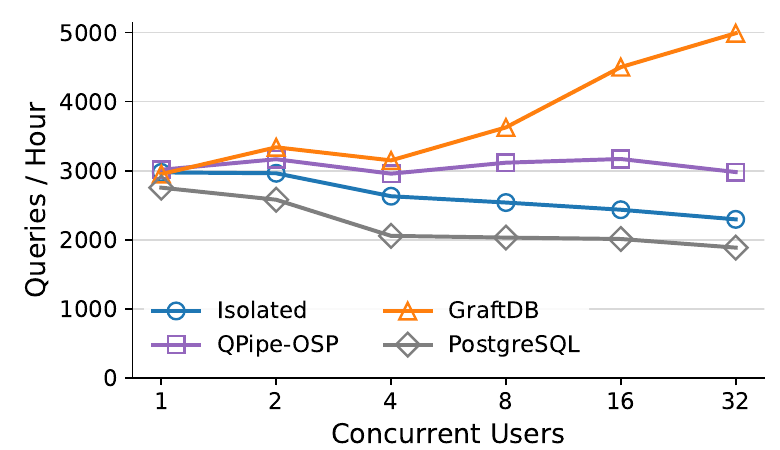}
  \caption{Closed-loop throughput as concurrency increases.
  \SYSNAME{} stays close to \emph{Isolated} at one client, then reaches 2.17 times higher
  throughput than \emph{Isolated} at 32 clients as more arrivals overlap with ongoing shared
  executions.}
  \Description{A line chart with concurrent users on the x-axis and query throughput on the y-axis.
  The chart compares Isolated, QPipe-OSP, GraftDB, and PostgreSQL over 1, 2, 4, 8, 16, and
  32 users.}
  \label{fig:eval-throughput-users}
\end{figure}

\begin{figure}[t]
  \centering
  \includegraphics[width=\linewidth]{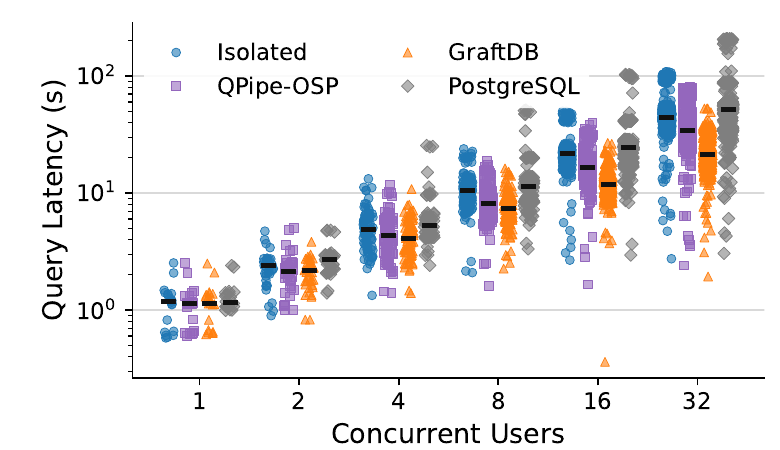}
  \caption{Workload-level query latency as closed-loop concurrency increases.
  Each point is one query execution, and thick marks show the median on a logarithmic y-axis.
  State-lens observations reduce repeated work at high concurrency, lowering median latency to
  0.48 times the \emph{Isolated} median at 32 clients.}
  \Description{A point plot with concurrent users on the x-axis and query latency on a logarithmic
  y-axis.
  The plot compares Isolated, QPipe-OSP, GraftDB, and PostgreSQL; points show individual query
  executions, and thick horizontal marks show medians.}
  \label{fig:eval-latency}
\end{figure}

\begin{figure*}[!t]
  \centering
  \includegraphics[width=\textwidth]{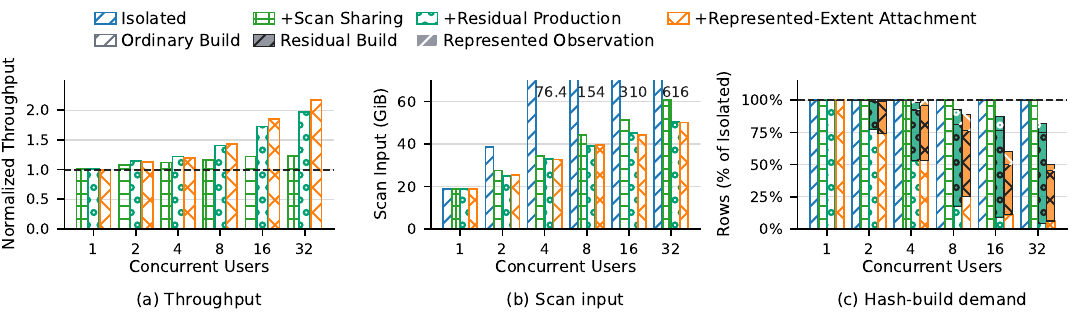}
  \caption{Mechanism breakdown on the closed-loop SF1 workload.
  Variants cumulatively enable scan sharing, residual production, and represented-extent
  attachment across throughput, scan input, and hash-build demand.
  (a) At 32 clients, the variants reach 1.23 times, 1.97 times, and
  2.17 times \emph{Isolated} throughput.
  (b) Scan sharing cuts scan input from 616\,GiB to 60.7\,GiB, and the two
  state-centric variants keep scan input near 50\,GiB.
  (c) Represented-extent attachment reduces exposed hash-build demand from
  82.3\% to 50.3\% of \emph{Isolated}.
  }
  \Description{Three-panel figure.
  Subfigure (a) is a bar chart with concurrent users on the x-axis and throughput
  normalized to Isolated on the y-axis.
  Subfigure (b) is a bar chart with concurrent users on the x-axis and scan input on
  the y-axis.
  Subfigure (c) is a stacked bar chart with concurrent users on the x-axis and
  hash-build rows normalized to Isolated on the y-axis.
  The visible stacks in (c) show ordinary builds, residual builds, and represented
  observations; unfilled space below 100 percent indicates hash-build demand bypassed
  by downstream state-lens observations.
  The subfigures compare scan sharing, residual production, and represented-extent attachment
  over 1, 2, 4, 8, 16, and 32 users; subfigures (b) and (c) also include Isolated.}
  \label{fig:eval-mechanism-breakdown}
\end{figure*}

\subsection{Closed-loop Concurrency}
\label{sec:evaluation-closed-loop}

The closed-loop run set uses TPC-H SF1 data and the default Zipf template
distribution over Q1 and Q3--Q10.
Each client executes 20 generated query instances and has at most one
outstanding query:
it submits the next query after its previous query completes.
We vary the number of clients over \{1, 2, 4, 8, 16, 32\}.
All systems use the same per-client query-instance sequences.
Figure~\ref{fig:eval-throughput-users} shows closed-loop throughput as
concurrency increases.

At one client, \SYSNAME{} stays close to \emph{Isolated} at 0.99 times its
throughput.
At 8, 16, and 32 clients, \SYSNAME{} achieves 1.43, 1.85, and 2.17 times
higher throughput than \emph{Isolated}.
In absolute terms, \SYSNAME{} rises from 3.0K to 5.0K queries/hour across the
sweep, while \emph{Isolated} falls from 3.0K to 2.3K and PostgreSQL from 2.8K
to 1.9K.
The improvement grows with concurrency because more query executions overlap
with running shared executions.

Figure~\ref{fig:eval-latency} plots every measured closed-loop query execution
and marks the scenario median.
At 32 clients, \SYSNAME{} reduces median latency from 44.4\,s to 21.3\,s, or
to 0.48 times the \emph{Isolated} median.
\emph{QPipe-OSP} reaches 34.5\,s at the same point.
At 16 clients, median latency falls from 21.6\,s to 11.9\,s.
The one-client median is similar across systems.

\subsection{Mechanism Breakdown}
\label{sec:evaluation-mechanisms}

The mechanism experiments build up \SYSNAME{} cumulatively on the same
closed-loop SF1 run set.
\emph{+Scan Sharing} shares base-table scans but does not expose shared operator
state to later-arriving queries.
\emph{+Residual Production} lets admitted producer paths contribute residual extents to
common shared state.
\emph{+Represented-Extent Attachment} additionally lets a later-arriving query observe a
compatible extent that the shared execution has already represented.
Figure~\ref{fig:eval-mechanism-breakdown} reports throughput, scan input, and
hash-build demand under this cumulative breakdown.

Figure~\ref{fig:eval-mechanism-breakdown}(a) and
Figure~\ref{fig:eval-mechanism-breakdown}(b) show that scan sharing removes most scan
input but not most of the throughput gain.
At 32 clients, \emph{+Scan Sharing} reduces scan input from 616\,GiB to 60.7\,GiB,
or 0.099 times \emph{Isolated}, and reaches 1.23 times \emph{Isolated} throughput.
The state-centric variants keep scan input low, ending at 0.081 times
\emph{Isolated}, while raising throughput to 1.97 times and 2.17 times
\emph{Isolated}.

Figure~\ref{fig:eval-mechanism-breakdown}(c) decomposes hash-build demand normalized to
the build-side rows that \emph{Isolated} would feed to hash-build state boundaries.
\emph{Ordinary Build} rows are unattached state-side input that remains ordinary-plan
work.
\emph{Residual Build} rows are residual extents inserted into shared state.
\emph{Represented Observation} rows are satisfied at the same boundary by observing a
represented extent through a state lens, rather than by inserting those rows on behalf of
the query.
When a bar does not reach 100\%, the unfilled portion represents isolated-plan hash-build
demand for upstream state that is no longer constructed for that query in the folded
execution after query grafting attaches the query at a downstream shared state.

At 32 clients, \emph{+Residual Production} leaves 82.3\% of \emph{Isolated}
hash-build demand exposed to hash-build state boundaries:
4.2\% ordinary builds, 71.1\% residual builds, and 7.0\% represented observations.
With \emph{+Represented-Extent Attachment}, the exposed demand falls to 50.3\%:
6.2\% ordinary builds, 39.0\% residual builds, and 5.1\% represented observations.
The remaining 49.7\% is upstream hash-build demand eliminated by downstream
state-lens observations.
Thus represented-extent attachment helps both directly, by observing represented extents,
and indirectly, by avoiding upstream hash-build work for attached queries.

\subsection{Open-loop Arrivals}
\label{sec:evaluation-open-loop}

\begin{figure}[t]
  \centering
  \includegraphics[width=\linewidth]{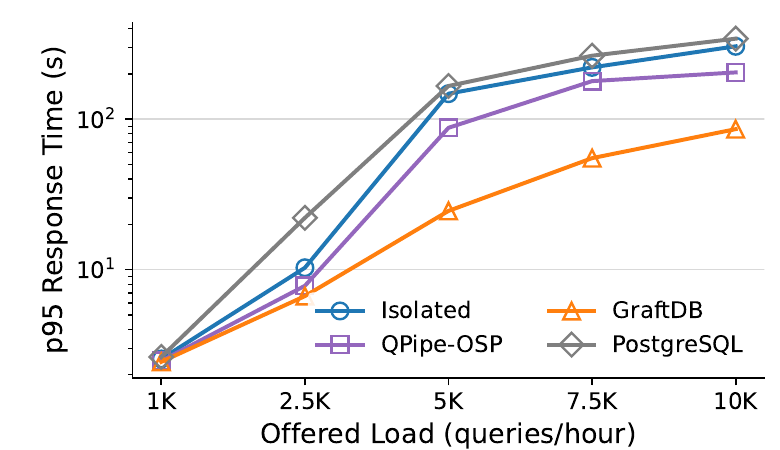}
  \caption{P95 response time under Poisson open-loop arrivals as offered load increases.
  \SYSNAME{} delays the queueing growth seen in the baselines, with the largest relative reduction
  at 5K offered queries/hour, where P95 response time falls to 0.17 times \emph{Isolated}'s P95
  response time.}
  \Description{A line chart with offered load on the x-axis and P95 response time on a logarithmic
  y-axis.
  The chart compares Isolated, QPipe-OSP, GraftDB, and PostgreSQL from 1000 to 10000 offered
  queries per hour.}
  \label{fig:eval-open-loop-response}
\end{figure}

Figure~\ref{fig:eval-open-loop-response} evaluates the same SF1 template set
under open-loop Poisson arrivals.
Each run first warms up the system for 120\,s at 1K queries/hour.
The measurement phase then submits queries for 60\,s at the offered load shown
on the x-axis.
After the measurement phase, no new queries are submitted, and the run waits
until all submitted queries complete.

For each offered load, scheduled arrival times are sampled from a Poisson
process and paired with query instances from the default template distribution.
The x-axis sets the mean arrival rate of that process, from 1K to 10K
queries/hour.
All systems replay the same scheduled arrival trace and query-instance sequence
for a given offered load.
Response time is measured from scheduled arrival time to query completion.

At 1K offered queries/hour, all systems keep P95 response time near 2.5\,s.
As offered load increases, \SYSNAME{} maintains lower P95 response time than
the baselines.
The largest relative reduction appears at 5K offered queries/hour, where P95
response time falls from 148.0\,s under \emph{Isolated} to 24.6\,s under
\SYSNAME{}, or to 0.17 times the \emph{Isolated} P95 response time.
\emph{QPipe-OSP} reaches 87.8\,s at the same point.
At 10K offered queries/hour, \SYSNAME{} reduces P95 response time from
305.7\,s to 86.0\,s, or to 0.28 times the \emph{Isolated} P95 response time,
while \emph{QPipe-OSP} reaches 205.0\,s.

\subsection{Sensitivity to Skew and Scale}
\label{sec:evaluation-sensitivity}

\begin{figure}[t]
  \centering
  \includegraphics[width=\linewidth]{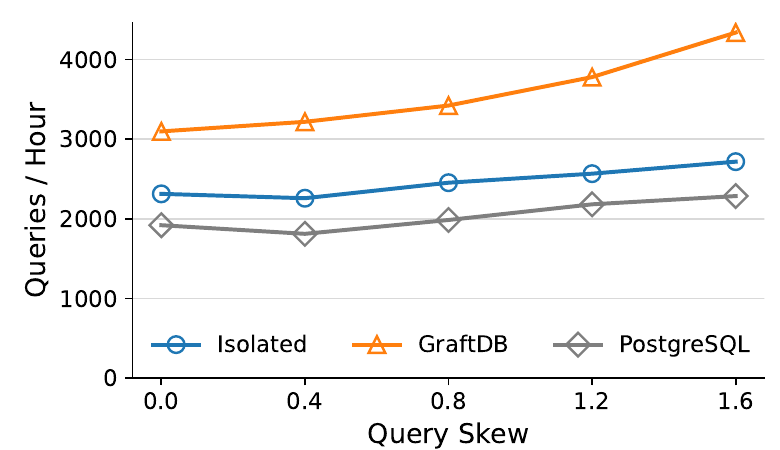}
  \caption{Throughput at fixed eight-client concurrency as query skew increases.
  Higher skew concentrates arrivals on fewer templates and increases overlap in operator
  requirements, raising \SYSNAME{} from 1.34 to 1.60 times higher throughput than
  \emph{Isolated}.}
  \Description{A line chart with query skew on the x-axis and query throughput on the y-axis.
  The chart compares Isolated, GraftDB, and PostgreSQL at fixed eight-user concurrency over Zipf
  skew values from 0.0 to 1.6.}
  \label{fig:eval-throughput-skew}
\end{figure}

Figure~\ref{fig:eval-throughput-skew} varies workload concentration by changing
the Zipf parameter of the template distribution from $\alpha = 0.0$ to
$\alpha = 1.6$ at fixed eight-client concurrency on the SF1 run set.
Each client follows the same 20-query closed-loop rule.
Parameter values within each template are still sampled uniformly from the same
large domains.

At $\alpha = 0.0$, which gives uniform template selection, \SYSNAME{} achieves
1.34 times higher throughput than \emph{Isolated}.
At $\alpha = 1.6$, the throughput ratio grows to 1.60 times.
Higher template skew concentrates arrivals on fewer templates, while the large
parameter domains keep exact duplicate query instances rare.
The resulting overlap is mainly overlap in operator requirements among related
but non-identical queries.

\begin{figure}[t]
  \centering
  \includegraphics[width=\linewidth]{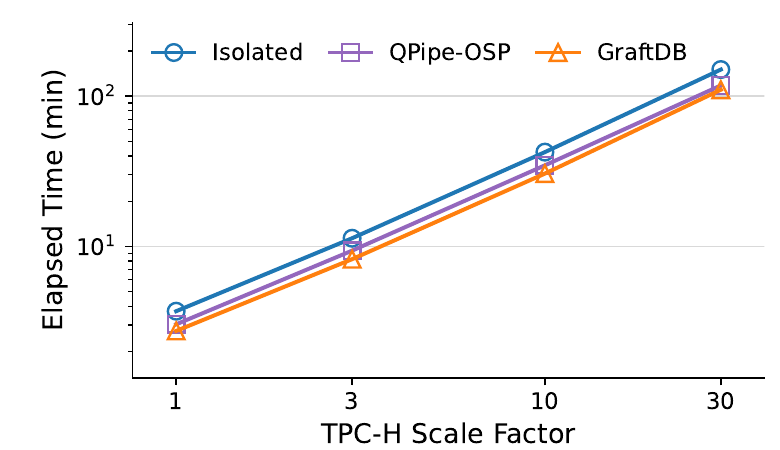}
  \caption{Elapsed time to complete the fixed eight-client workload as TPC-H scale factor
  increases.
  \SYSNAME{} remains faster than both baselines across SF1--SF30, staying between 0.72 and
  0.74 times the \emph{Isolated} completion time.}
  \Description{A line chart with TPC-H scale factor on a logarithmic x-axis and elapsed
  completion time on a logarithmic y-axis.
  The chart compares Isolated, QPipe-OSP, and GraftDB over scale factors 1, 3, 10, and 30.}
  \label{fig:eval-scale-factor}
\end{figure}

Figure~\ref{fig:eval-scale-factor} varies data scale using the same eight-client
closed-loop workload shape at TPC-H SF1, SF3, SF10, and SF30.
It reports workload completion time on logarithmic axes.
\SYSNAME{} completes faster than \emph{Isolated} throughout the sweep.
At SF30, completion time falls from 150.5 minutes under \emph{Isolated} to
110.4 minutes under \SYSNAME{}.
\emph{QPipe-OSP} completes the same workload in 117.9 minutes.
Across SF1--SF30, \SYSNAME{} stays between 0.72 and 0.74 times the
\emph{Isolated} completion time.

\section{Related Work}
\label{sec:related-work}

\SYSNAME{} shares analytical work at the arrival-time boundary of a dynamic
concurrent workload.
When a query arrives, the runtime can attach it to operator state that a running
execution has already produced, while producer work that can still receive source input
can contribute to that state.
This decision differs from sharing over a saved artifact, a preselected query
set, a running scan or pipeline, or maintained streaming state.
The shared object is partially produced operator state; the admission test asks
whether the represented extent can be observed through a per-query state lens and whether
remaining source input can still feed producer work that contributes the residual extent
to the same state.

Result reuse and materialized-view systems make completed work available to later
queries.
A query can use a materialized view, recycled intermediate, or MapReduce job
result when that artifact matches the query or can answer
it~\cite{conf/pods/LevyMSS95,conf/icde/ChaudhuriKPS95,
conf/sigmod/GoldsteinL01,conf/sigmod/KotidisR99,conf/sigmod/IvanovaKNG09,
conf/icde/NagelBV13,conf/sigmod/MistryRSR01,journals/ftdb/ChirkovaY12,
journals/pvldb/ElghandourA12}.
Hybrid MQO work combines materialized-view reuse with shared subexpression reuse
in a batched optimization setting~\cite{journals/isf/GurumurthyBBPS24}.
Semantic caching, internal data-structure reuse, reactive caching, and
intermittent query processing preserve reusable extents, structures, or operator
state across related computations~\cite{conf/vldb/DarFJST96,
conf/sigmod/DursunBCK17,journals/pvldb/AzimKA17,journals/pvldb/TangSEKF19}.
\SYSNAME{} instead keeps partially produced state inside a running shared
execution and controls each query's observation through a per-query state lens.
Multi-query optimization and batched shared-execution systems move the decision
earlier:
they reason over a known or admitted query set and construct shared plans,
subplans, or execution cycles before that shared unit
runs~\cite{conf/sigmod/Finkelstein82,journals/tods/Sellis88,
conf/sigmod/RoySSB00,conf/pods/DalviSRS01,conf/sigmod/ZhouLFL07,
conf/bigdataconf/TuEXC22,journals/pvldb/NykielPMKK10,journals/pvldb/WangC13,
journals/pvldb/El-HelwRSCGP15,
journals/pvldb/GiannikisAK12,journals/pvldb/GiannikisMAK14,
journals/pvldb/MakreshanskiGAK16}.
\SYSNAME{} uses a different boundary.
It admits a query after execution has already changed operator state, then
partitions the query's state-side extent into represented, residual, and unattached
extents.

Runtime-sharing systems make sharing decisions while execution is in progress.
Shared scans coordinate concurrent access to base
tables~\cite{conf/vldb/ZukowskiHNB07,journals/pvldb/QiaoRRHL08,
conf/icde/RamanSQRDKNS08}, QPipe shares operator pipelines across concurrent
queries~\cite{conf/sigmod/HarizopoulosSA05}, and CJoin, Crescando, DataPath,
and studies of concurrent analytical work sharing coordinate joins,
data-centric paths, or global query plans under changing
workloads~\cite{journals/pvldb/CandeaPV09,journals/pvldb/UnterbrunnerGAFK09,
conf/sigmod/ArumugamDJPP10,journals/pvldb/PsaroudakisAA13,
conf/sigmod/PsaroudakisAOA14}.
These systems share running work through scan policies, operator pipelines, tuple
routing, or shared join structures.
Hash teams similarly organize join and group-by execution around hash
structures inside a plan~\cite{conf/vldb/KemperKW99}.
\SYSNAME{} also makes an online decision, but the decision is made at a stateful
boundary after earlier work has produced state.
A folded hash-probe step is admitted only when coverage metadata describes the relevant
build-side extent as complete, source availability can contribute the residual build-side
extent to the selected state, and the data-flow path can supply the probe-side input.

Adaptive query processing also exposes or manipulates execution state while a
query is running.
SteMs decompose join processing into state modules, and STAIRs make join state
explicitly modifiable and migratable during adaptive execution~\cite{
conf/icde/RamanDH03,conf/vldb/DeshpandeH04}.
Tukwila, proactive re-optimization, progressive optimization, and AQP surveys
use runtime feedback or adaptive routing to change execution
decisions~\cite{conf/sigmod/IvesFFLW99,conf/sigmod/BabuBD05,
conf/sigmod/MarklRSLP04,conf/cidr/BabuB05,journals/ftdb/DeshpandeIR07}.
\SYSNAME{} instead uses partially produced operator state as an inter-query
sharing unit for later-arriving finite analytical queries.

Stream and continuous-query systems maintain computation across changing inputs
or changing query sets.
STREAM, CQL, Aurora, and Borealis define semantics or architectures for data
stream management and continuous queries~\cite{conf/cidr/MotwaniWABBDMORV03,
journals/vldb/ArasuBW06,journals/vldb/AbadiCCCCLSTZ03,
conf/cidr/AbadiABCCHLMRRTXZ05}.
CACQ~\cite{conf/sigmod/MaddenSHR02},
TelegraphCQ~\cite{conf/cidr/ChandrasekaranDFHHKMRRS03}, and
PSoup~\cite{conf/vldb/ChandrasekaranF02} provide adaptive or shared
continuous-query processing structures, and AStream and AJoin support ad-hoc
stream queries that can be created and deleted while the system
runs~\cite{conf/sigmod/KarimovRM19,journals/pvldb/KarimovRM19}.
Flux repartitions continuous-query state, and MJoin optimizes multi-way stream
joins~\cite{conf/icde/ShahHCF03,conf/vldb/ViglasNB03}.
Shared arrangements make indexed arrangement state available across concurrent
streaming dataflows~\cite{journals/pvldb/McSherryLSR20}, and timely dataflow
and recent stream-join systems optimize progress tracking, join models, or join
orders under streaming updates~\cite{conf/sosp/MurrayMIIBA13,
journals/pacmmod/WangZZSLH24,journals/pacmmod/YeGZZ25}.
These systems make state shareable under stream time, windows, maintained
versions, or continuous updates.
\SYSNAME{} targets finite analytical executions.
For hash-build state, coverage metadata describes which finite build-side extent the shared
state represents; for aggregate state, exact aggregate identity determines which
completed state can be observed.
These conditions let a later one-shot query observe only the state-side extent that is
complete for that query's state lens and wait only for residual producer work admitted
to the same state.

Other systems optimize adjacent execution boundaries.
DynQ reuses compiled-query artifacts in a polyglot
runtime~\cite{journals/vldb/SchiavioBB23}, and Lemo uses cached subquery and
intermediate results in learned optimization for concurrent
queries~\cite{journals/pacmmod/MoCWCB23}.
DBToaster, Differential Dataflow, and Noria maintain results or dataflow state as
inputs change, with Noria also sharing state across related application
queries~\cite{journals/vldb/KochAKNNLS14,conf/cidr/McSherryMII13,
conf/osdi/GjengsetSBAEKKM18}.
Pipeline group optimization, sideways information passing, predicate transfer,
Eddies, and NiagaraCQ improve execution through pipeline placement, prefiltering,
adaptive routing, or dynamic continuous-query grouping~\cite{conf/cidr/GeyerKHL23,
conf/icde/IvesT08,conf/cidr/YangZYK24,conf/sigmod/HellersteinA00,
conf/sigmod/ChenJDTW00}.
\SYSNAME{} instead makes partially produced operator state the unit of dynamic
folding for analytical queries that arrive over time.
Per-query state lenses restrict what each query may observe from that state, and
query grafting turns the compatible part of the arriving query into shared-execution
work.

\section{Conclusion}
\label{sec:conclusion}

\SYSNAME{} shows that dynamic concurrent analytical workloads can share work through
operator state accumulated by a running execution.
Hash-build and aggregate states become shared objects that later-arriving queries can
attach to while preserving per-query semantics.
Per-query state lenses define what each query may observe, while query grafting
assigns state-side input to represented, residual, and unattached extents.

Experiments on TPC-H-derived dynamic concurrent workloads show that this
execution model reduces redundant work across overlapping analytical queries.
\SYSNAME{} achieves up to 2.17 times higher throughput than the isolated baseline
and, under overloaded open-loop arrivals, reduces P95 response
time to as low as 0.17 times the isolated baseline's P95 response time.
The mechanism breakdown shows that scan sharing alone does not explain these gains:
residual production and represented-extent attachment reduce hash-build demand after
scan input has largely been removed.

The result is a runtime sharing unit for dynamic concurrent workloads: partially
accumulated operator state.
Future work will extend dynamic folding to multi-threaded execution.
This extension must coordinate concurrent state producers and per-query
state-lens observations while preserving the completeness and visibility
conditions used by the single-threaded prototype.

\begin{acks}
This work was supported in part by JSPS Grant-in-Aid for Research Fellows JP24KJ0769
and JSPS Grant-in-Aid for Scientific Research (B) JP26K02915.
\end{acks}

\bibliographystyle{ACM-Reference-Format}
\bibliography{references}

\end{document}